%% file: bcTsneDIFF.tex
\documentclass[11 pt]{article}

\input{pack.tex}

\usepackage{natbib}
\usepackage{authblk}
\newcommand{\x}{\boldsymbol{x}}
\newcommand{\y}{\boldsymbol{y}}

\newcommand{\X}{\textbf{X}}
\newcommand{\Y}{\textbf{Y}}
\newcommand{\Z}{\textbf{Z}}
\newcommand{\tr}{^{\mkern-1.5mu\mathsf{T}}}
\DeclareMathOperator*{\argmin}{argmin}

\begin{document}

\title{Projected $t$-SNE for batch correction}
\author[1]{Emanuele Aliverti}
\author[2]{Jeff Tilson}
\author[2,3]{Dayne Filer}
\author[3,4]{Benjamin Babcock}
\author[3]{Alejandro Colaneri}
\author[4,5]{Jennifer Ocasio}
\author[4,5,6,7]{Timothy R. Gershon}
\author[2,3,4]{Kirk C. Wilhelmsen}
\author[8]{David B. Dunson}

\small{
\affil[1]{Department of Statistical Sciences, University of Padova, Padova}
\affil[2]{RENCI, Chapel Hill, NC 27517 USA}
\affil[3]{Department of Genetics, University of North Carolina School of Medicine, Chapel Hill}
\affil[4]{Department of Neurology, University of North Carolina School of Medicine, Chapel Hill}
\affil[5]{UNC Neuroscience Center, University of North Carolina School of Medicine, Chapel Hill}
\affil[6]{Carolina Institute for Developmental Disabilities, University of North Carolina School of Medicine, Chapel Hill}
\affil[7]{Lineberger Comprehensive Cancer Center, University of North Carolina School of Medicine, Chapel Hill}
\affil[8]{Department of Statistical Science, Duke University, Durham}
}

\renewcommand*{\Authfont}{\normalsize}
\renewcommand*{\Affilfont}{\small\normalfont}
\date{}
\maketitle

\abstract
{\textbf{Motivation:} 
	Low-dimensional representations of high-dimensional data are routinely employed in biomedical research to visualize, interpret and communicate results from different pipelines.
	In this article, we propose a novel procedure to directly estimate $t$-SNE embeddings that are not driven by batch effects. Without correction, interesting structure in the data can be obscured by batch effects.  The proposed algorithm can therefore significantly aid visualization of high-dimensional data.
\\
\textbf{Results:} 
The proposed methods are based on linear algebra and constrained optimization, leading to efficient algorithms and fast computation in many high-dimensional settings.
Results on artificial single-cell transcription profiling data show that the proposed procedure successfully removes multiple batch effects from $t$-SNE embeddings, while retaining fundamental information on cell types.
When applied to single-cell gene expression data to investigate mouse medulloblastoma, the proposed method successfully removes batches related with mice identifiers and the date of the experiment, while preserving clusters of oligodendrocytes, astrocytes, and endothelial cells and microglia, which are expected to lie in the stroma within or adjacent to the tumors. \\
\textbf{Availability:} Source code implementing the proposed approach in R and Julia is available at the link \href{https://github.com/emanuelealiverti/BC_tSNE}{https://github.com/emanuelealiverti/BC\_tSNE}, including a tutorial to reproduce the simulation studies.\\
\textbf{Contact:} \href{aliverti@stat.unipd.it}{aliverti@stat.unipd.it}\\
\textbf{Supplementary information:} Supplementary data are available at \textit{Bioinformatics} online.}

\maketitle

\section{Introduction}
Recent technological improvements in transcriptome analysis have lead to many valuable insights into complex biological systems, with single-cell RNA transcription profiling (scRNAseq) analysis being one of the most popular tools to investigate intricate cellular processes \citep{hwang2018single}.
In biostatistical analysis, low-dimensional representations of high-dimensional scRNAseq data are ubiquitous, playing a central role during multiple phases of scientific investigation.
For example, visualisation tools are used during normalisation, correction and dimensionality reduction to evaluate success of the pipelines, and in downstream analysis to illustrate results from intermediate procedures such as clustering \citep[e.g.][]{luecken2019current,lun2016step,vieth2019systematic}.

A wide variety of methods for linear and non-linear dimensionality reduction and data visualisation are available, with $t$-distributed Stochastic Neighbor Embedding \cite[$t$-SNE,][]{maaten2008visualizing} and Uniform Manifold Approximation and Projection \cite[UMAP,][]{mcinnes2018umap} being of great utility in analyzing scRNAseq data.
Such methods allow one to describe the dataset in 2-3 dimensions via graphical representations, highlighting the main structure of the data and preserving relevant properties such as the presence of isolated clusters \citep{kobak2019art}.
During pre-processing, low-dimensional representations are fundamental for identifying potential issues in the data; for example, inadequate data integration or the presence of batch effects \citep{luecken2019current,lun2016step}.
Indeed, without explicit adjustment, variations in the low-dimensional summaries may be driven by nuisance covariates --- such as batches due to different devices used for an experiment --- instead of the primary factors of scientific interest --- such as cell types.  In intermediate analyses, such batch effects can limit the utility of the low-dimensional graphical representations in visualizing, interpreting and communicating results from downstream processes conducted at the cell level; for example, clustering, cell annotations or compositional analysis \citep{wagner2016revealing}.

In a typical workflow, standardized pipelines proceed sequentially, with low-dimensional embeddings estimated after several steps involving normalisation, integration, batch-correction and feature selection from raw data; see \citet{luecken2019current} and references therein for a recent detailed review.
However, such processing might lead to propagation of errors and unreliable representations. 
For example, over-correction of batch effects might also remove important biological features, and lead to low-dimensional embeddings which are not driven by such biological factors \citep[e.g.][]{lun2016step}.
Such an issue will be entirely propagated to downstream processes, leading to low-dimensional embeddings which cannot highlight information on factors of interest and might provide misleading evidence. 

Motivated by the above considerations, the focus of this article is on producing batch-corrected modifications of $t$-SNE that can be used to remove associations with multiple batches from low-dimensional embeddings.
Our methods are based on linear algebra results and modification of gradient descent optimisation, therefore providing simple and scalable tools in high-dimensional problems.
The proposed procedure directly estimates low-dimensional embeddings, which are not driven by systematic batch-effects including batch-correction,  and provides a synthetic representation to correctly visualise results from different pipelines.

Several approaches are available in the literature for batch-correction and data integration, covering a wide range of methods which encompass linear modelling via Empirical-Bayes \citep{johnson2007adjusting}, canonical correlation analysis \citep{butler2018integrating} and Mutual Nearest Neighbors \citep[MNN,][]{haghverdi2018batch}; see \citet{buttner2019test} and references therein for a recent comparison, and the \texttt{scater} package \citep{mccarthy2017scater} for a convenient implementation.
Differently from routine corrections for scRNAseq data, our approach is not targeted to correct the entire set of high-dimensional data, but only its low-dimensional representation obtained via $t$-SNE.
Therefore, the proposed approach directly relates to the framework of ``removal of unwanted variations'' \citep[RUV. See ][]{grun2015design,risso2014normalization,leek2007capturing}, where interest is on measuring latent variables which are not affected by batch-effects and experimental conditions, but are only driven by relevant biological factors.

Specifically, we introduce a novel modification of $t$-SNE to integrate batch correction into estimation of low-dimensional embeddings. 
Such an approach is not intended as a substitute to the canonical pipelines for downstream analysis; which, for example, focus on estimating clusters in the $k$-NN graph of the PCA subspace \citep[e.g.][]{wolf2019paga}.
Instead, the proposed contribution serves as a parallel tool to provide a robust visualisation of scRNAseq data, which is less subject to propagation of errors and can be used to validate results from different pipelines, or to identify potential pitfalls.
Although there is some evidence that clustering in the $t$-SNE subspace can provide insights on the community structure of the data \citep{linderman2019clustering}, such a procedure is generally not recommended in the analysis of scRNAseq data and is beyond the scope of the current article; see \citet{kobak2019art} for further discussion.
The proposed algorithm allows joint correction for multiple batches and leverages two different adjustments, to handle both linear and non-linear effects.
Linear correction is achieved adapting the strategy of \citet{aliverti2018removing}, while correction for non-linear effects leverages a projection step during $t$-SNE optimisation, related to the locally linear correction implemented in \citet{haghverdi2018batch}.
A full implementation of the method is publicly available in a mixed R and Julia implementation, at the link \href{https://github.com/emanuelealiverti/BC_tSNE}{\texttt{https://github.com/emanuelealiverti/BC\_tSNE}}.

\section{Methods}

\subsection{Notation \& problem formulation}
Consider a data matrix ${\X} = \{\x_i\tr\}_{i=1}^n$ with observations ${\x_i = (x_{i1}, \dots, x_{ip})}$. 
In many biological applications, the number of features $p$ is tremendously large and it is of interest to provide accurate low-dimensional representation of such high dimensional data.
Dimensionality reduction techniques focus on finding low-dimensional counterparts ${\y_i = (y_{i1}, \dots, y_{iq})}$ of each $\x_i$, preserving as much structure as possible with $q \ll p$ components; generally, $q=2$ or $q=3$ for the ease of graphical visualisation.
Original observations $\x_i$ can potentially lie in a complex and highly non-linear manifold; for example, wrapped spaces such as rolls \cite[e.g][]{lee2005nonlinear}.
In contrast, the desired low-dimensional embedding lie on a standard $q$-dimensional Euclidean space, and $\y_i$ determines the position of observation $i$ in such an embedded space.

Low-dimensional representations are constructed in order to preserve some specific structure of the original data; some examples include preserving Euclidean distances \citep[Multidimensional Scaling, ][]{kruskal1964multidimensional}, variances (Principal Component Analysis), neighborhood graphs \cite[Local Linear Embedding and Isomap,][]{roweis2000nonlinear,tenenbaum2000global} or local similarities among points \citep[Stochastic Neighbor Embedding,][]{hinton2003stochastic}.
Many methods estimate an explicit function between the original data and their embeddings; for example, the PCA solution is a linear combination of the columns of $\X$.
More recently, focus has shifted to obtaining $\y_i$ without explicitly defining such a map, thus allowing a greater flexibility and range of application.
In this article, we focus on the $t$-SNE methodology for dimensionality reduction and data visualisation \citep{maaten2008visualizing}. 
$t$-SNE attempts to find low-dimensional representations that preserve local similarities among data points, with similarities parameterized as conditional probabilities of belonging to the same local neighborhood. In the following paragraphs, we review the standard formulation of $t$-SNE before introducing our adjustments for batch effects. 

\subsection{Standard $t$-SNE algorithm}
In the original input space, $t$-SNE defines dissimilarities among points as symmetric probabilities $p_{ij} = (p_{i\mid j} + p_{j \mid i}) / 2n $, with
 \begin{equation}
	 \label{eq:condProb}
	 p_{i\mid j} = \frac{\exp\left(-0.5||\x_i - \x_j||^2 / \sigma_i^2\right)}{\sum_{k, k\neq i}\exp\left(-0.5||\x_i - \x_k||^2 / \sigma_i^2\right)}.
 \end{equation}
Equation~\ref{eq:condProb} can be interpreted as the probability that point $i$ picks $j$ as its neighbor, under a Gaussian kernel centered at $\x_i$ and with standard deviation equal to $\sigma_i$.
The intuition behind the introduction of $p_{ij}$ comes from averaging $p_{i\mid j}$ and $p_{j \mid i}$ to reduce the relative impact of outliers and define a symmetric dissimilarity metric \citep{maaten2008visualizing}.
The parameter $\sigma^2_i$ determines the width of the Gaussian kernel and, indirectly, the number of local neighbors associated with each point $i$, with $i=1,\dots,n$.
Defining $\sigma^2_i$ is a primary step in producing $t$-SNE embeddings, with the selection determining the \emph{perplexity} of the resulting distribution \citep{maaten2008visualizing,hinton2003stochastic}.
Large values of $\sigma^2_i$ correspond to a larger number of local neighbors and greater perplexity, while default values of perplexity range in the interval $\{10,50\}$ \citep{maaten2008visualizing}.
Embeddings generally show robustness to moderate changes in perplexity \citep{van2014accelerating}.

 Dissimilarity among points $\y_i$ in the embedded space is defined through the kernel of a $t$-distribution with $1$ degree of freedom, setting
 \begin{equation}
	 \label{eq:tProb}
	 q_{ij} = \frac{\left(1+||\y_i - \y_j||^2 \right)^{-1}}{\sum_{k, k\neq i}  \left(1+||\y_i - \y_k||^2 \right)^{-1}}.
 \end{equation}
The $t$-SNE embeddings $\y_i$ are selected minimizing the Kullback-Leibler divergence between $p_{ij}$ and $q_{ij}$; note that $p_{ij}$ does not depend on $\y$ and is a fixed value given the input data.
Let $\y = (\y_1, \dots, \y_n)$ and highlight the dependency of $q_{ij}$ on the embeddings $\y$ in Equation~\ref{eq:tProb} as $q_{ij}(\y)$. 
Formally, $t$-SNE is the solution to the following optimisation problem.
 \begin{equation}
	 \label{eq:KL2}
	 \begin{aligned}
		 \argmin_{\y} \left\{L(\y)\right\} = \argmin_{\y} =  \left\{ \sum_{i=2}^n\sum_{j=1}^i p_{ij} \log \frac{p_{ij}}{q_{ij}(\y)} \right\}
	 \end{aligned}
 \end{equation}  
The objective function $L(\y)$ can be optimized through gradient methods. Indeed, the partial derivative of the loss functions in Equation~\ref{eq:KL2} with respect to $\y_i$ is equal to
 \begin{equation}
	 \label{eq:grad}
	 \begin{aligned}
\nabla L(\y_i) =  \frac{\partial L}{\partial \y_i} =  4 \sum_{\substack{j=1 \\ j\neq i} }^n (p_{ij} - q_{ij})q_{ij} Z(\y_i - \y_j); \\
	 Z =  \sum_{l\neq k} \big(1-||\y_l - \y_k||^2\big)^{-1};
	 \end{aligned}
 \end{equation}  
see \citet[][Appendix A]{maaten2008visualizing} for the complete derivation.
 
Therefore, the generic gradient descent step with momentum correction for updating $\y_i$ at iteration $t+1$ sets
\begin{equation}
	 \label{eq:gradU}
	\y_i^{(t+1)} = \y_i^{(t)} + \eta^{(t)} \nabla L(\y_i^{(t)}) +  \alpha^{(t)} \left(\y_i^{(t)} - \y_i^{(t-1)}\right),
 \end{equation}
 with $\eta^{(t)}$ indicating the learning rate and $\alpha^{(t)}$ the momentum term; see \citet{maaten2008visualizing} for practical advices on the choice of such functions.

\subsection{Batch-corrected $t$-SNE}
\label{sec:pp}
Let $\Z$ denote an additional variable which contains batch information.
We refer to the proposed method as BC-$t$-SNE (Batch-Corrected $t$-SNE) in the sequel.
When the number of features $p$ is extremely large and when it exceeds the number of observations $n$,
direct application of $t$-SNE on the raw data ${\X}$ can be challenging, computationally inefficient, and lead to poor results. 
Therefore, it is generally advised to perform a preliminary dimensionality reduction, and then apply $t$-SNE over such reduced representation to improve the results \citep{van2014accelerating}.
For example, default software implementation estimates $t$-SNE embeddings on the first $k$ principal components, with $k$ in the range $[30-50]$  \citep[e.g.][]{krijthe2018package}.
Reducing the dimensionality from $p$ to $k$ speeds up computation and reduces noise without affecting local similarities among observations \citep{maaten2008visualizing}. 

The first step of BC-$t$-SNE is motivated by the above considerations, and focuses on processing the data with the approach introduced in \citet{aliverti2018removing} to explicitly obtain the optimal low-rank approximation of a matrix $\X$ in Frobenius norm under an orthogonality constraint between such approximation and the batch variable $\Z$.
Therefore, the method removes linear effects between the reduced data and the variables in $\Z$  with minimal information loss.
Such an approach is based on computing the residuals from a multivariate regression among the left singular vectors of ${\X}$ and $\Z$, and is therefore comparable with standard PCA in terms of computational requirements, providing a practical alternative to perform dimensionality reduction while simultaneously achieving batch removal.
Although such a procedure is optimal in removing the linear influence of batches, effects beyond linearity might still affect $t$-SNE embeddings.
In practical applications such higher order effects are often small in magnitude, and second order adjustment often lead to satisfactory results in terms of batch-correction \citep[e.g.][]{aliverti2018removing}.
However, since the $t$-SNE embeddings $\y_i$ are a complex non-linear functional of the original $\x_i$, inclusion of higher-order constraints provides a reasonable conservative choice.

The second step of BC-$t$-SNE adjustment can be better motivated by introducing some details on gradient descent, which can be interpreted as an optimisation to minimize the linearisation of the loss function (Kullback-Leibler divergence for $t$-SNE) around the current estimates, including a smoothing penalty that penalises abrupt changes \citep[e.g.][]{Sparsity:2015}.
To see that, consider the gradient descent step in  Equation~\ref{eq:gradU}, setting without loss of generality $\alpha^{(t)} = 0$. The following alternative representation holds.
\begin{equation}
\begin{split}
	\y_i^{(t+1)} = \argmin_{\y_i \in \mathbb{R}^q} \bigg\{L(\y_i^{(t)}) +  \langle \nabla L(\y_i^{(t)}),\y_i - \y_i^{(t)}\rangle - \\ \frac{1}{\eta^{(t)}} ||\y_i - \y_i^{(t)}||^2_2  \bigg\}.
\end{split}
\end{equation}

This view of gradient descent facilitates the introduction of further constraints.
Indeed, this aim is achieved by restricting the solution ${\y_i \in \mathcal{C}}$, with $\mathcal{C}$ denoting a constrained region of the original space $\mathbb{R}^q$.
Such a constraint can be easily imposed by performing a standard gradient step, and then projecting the result back into the constrained set $\mathcal{C}$, leading to a procedure referred to as \emph{projected} gradient descent; see, for example, \citet[Sec 5.3.2]{Sparsity:2015} for further details. 
The choice of the constrained set $\mathcal{C}$ covers a central role in the optimisation, since the projection should be computed easily in order to make the method practical in high-dimensional applications.  
With this motivation in mind, we propose a computationally simple solution and restrict $\Y=\{\y_i^\intercal\}_{i=1}^{n}$ such that it is orthogonal with the subspace spanned by the columns of $\Z$.
This constraint can be easily imposed with linear regression, computing at each iteration a projected gradient step which constructs an update $\tilde{\Y}^{(t+1)}$ that projects the unconstrained solution  ${\Y}^{(t+1)}$, making it orthogonal with the batch variables $\Z$.
Pseudo-code illustrating the method is reported in Algorithm~\ref{alg:og}.

\section{Simulation study}
A simulation study is conducted to evaluate the performance of the proposed method on artificial scRNAseq data.
Artificial single-cell RNA sequencing data were generated with the BioConductor library \texttt{splatter}, which provides an interface to create complex datasets with several realistic features \citep{zappia2018splatter}.
Specifically, a dataset consisting of ${p=10000}$ genes measured over $n=800$ cells was generated with $4$ batch effects and $4$ different cell types.
A complete tutorial to reproduce the artificial data and simulation study in R and julia is available at the link \href{https://github.com/emanuelealiverti/BC_tSNE}{\texttt{https://github.com/emanuelealiverti/BC\_tSNE}}.

\RestyleAlgo{boxruled}
\begin{algorithm}[tb]
\caption{Batch-corrected-$t$-SNE with projected gradient}
\label{alg:og}
	Apply OG \citep{aliverti2018removing} to extract the first $k$ components of ${\X}$ and remove linear batch effects. Denote the $n\times k$ reduced and adjusted matrix as $\hat\X$
\For{$i = 1, \dots, n$} {
    	Perform binary search to find the value $\sigma_i^2$ that achieves desired level of \emph{perplexity} \citep{maaten2008visualizing}
}
	Compute the pairwise similarities $p_{ij}$ in Equation~\ref{eq:condProb} from $\hat\X$ and $\{\sigma_i^2\}_{i=1}^n$
    \For{$t=1,\dots,T$}{
    Compute affinities $q_{ij}$ defined in Equation~\ref{eq:tProb}
            \For{$i=1,\dots,n$}{
	    Update $\y_i^{{(t+1)}}$ (the $i$-th row of $\Y^{(t+1)}$) as $$\y_i^{(t+1)} = \y_i^{(t)} + \eta^{(t)} \nabla L(\y_i^{(t)}) +  \alpha^{(t)} \left(\y_i^{(t)} - \y_i^{(t-1)}\right)$$
    }
	Compute $\boldsymbol{\beta}^{(t+1)}\gets\text{solve}(\Z^\intercal\Z,\Z^\intercal \Y^{(t+1)})$
	Compute projected gradient update, setting $$\tilde{\Y}^{(t+1)} ={\Y}^{(t+1)} -\Z\boldsymbol{\beta^{(t+1)}}$$
}
	\KwOut{Return ${\tilde{\Y}^{(T)}}$.}
\end{algorithm}

The focus of the simulation is on assessing the success of BC-$t$-SNE at removing unwanted associations while retaining information of the scientific factors of interest, which correspond to cell types in this particular example.
The adjusted approach is also compared with a standard implementation of $t$-SNE, available with R package Rtsne \citep{rtsne}, and with routine methods for batch-correction.
In particular, we apply the recently proposed MNN \citep{haghverdi2018batch} and Harmony \citep{korsunsky2019fast} methods for batch-correction, available through the R packages \texttt{batchelor} and \texttt{harmony}.
In order to properly compare the methods, parameters of BC-$t$-SNE were fixed to the default configuration of the package Rtsne, which corresponds to setting the number of iterations $T=1000$, a value of perplexity equal to $30$ and $\eta^{(t)}=200, t = 1, \dots T$ and $\alpha^{(t)} = 0.5$ for $t<250$ and  $\alpha^{(t)} = 0.8$ for $t\geq 250$; see also \citet{maaten2008visualizing}.

\begin{figure}[t]
	\centerline{\includegraphics[width = .4\textwidth]{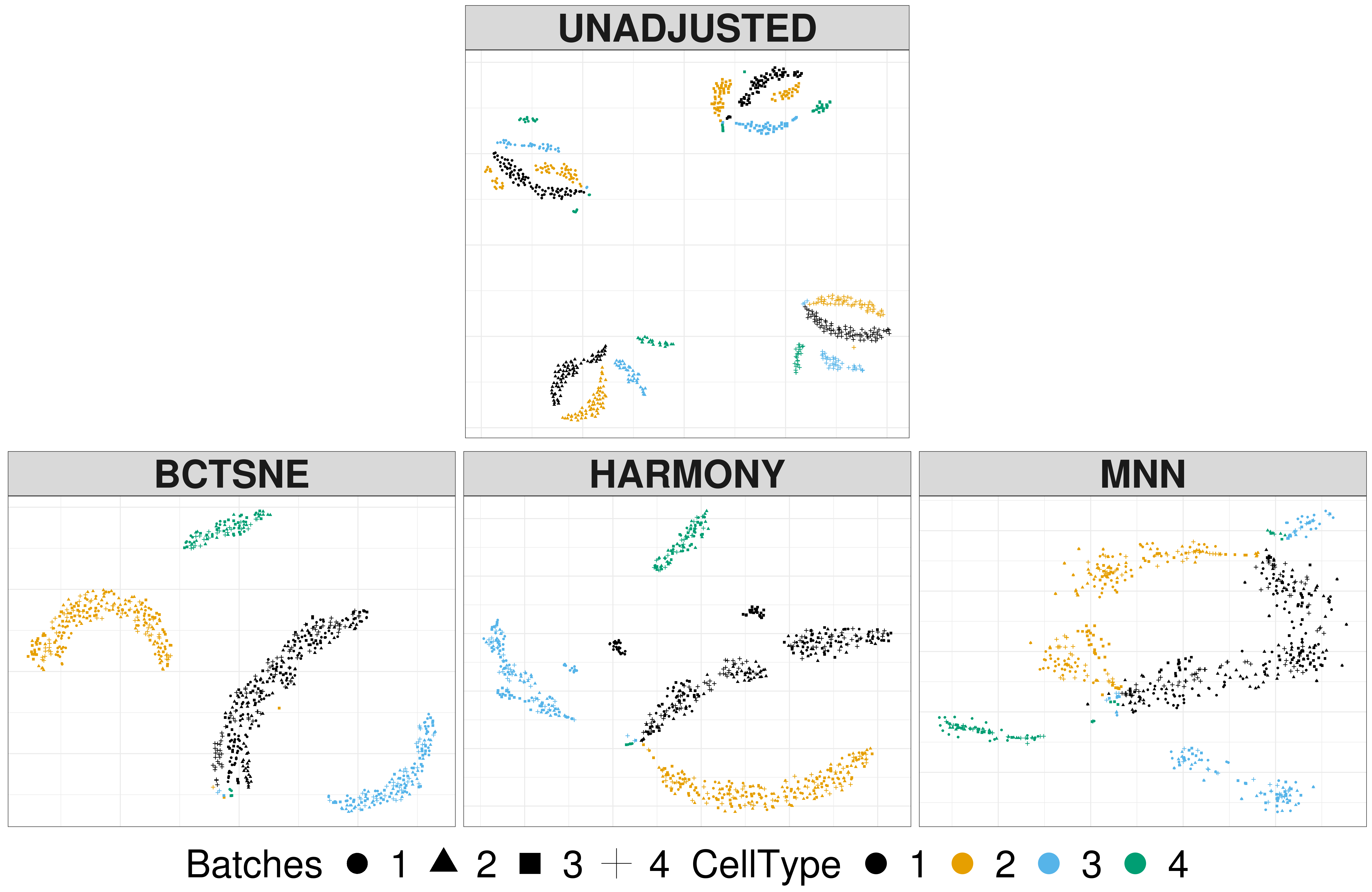}}
\caption{Simulation study. The color of points varies according to cell types, while shapes vary with batch groups. Upper plot shows the unadjusted $t$-SNE coordinates, while results after adjustment are reported in the bottom panels.}
\label{fig:sim1}
\end{figure}

Figure~\ref{fig:sim1} compares results from unadjusted $t$-SNE and the proposed method, respectively in the upper and lower panels; both approaches are estimated over a $k=30$ reduced components.
Results for the unadjusted case confirm the presence of strong batch effects.
Indeed, cells are divided into $4$ main clusters corresponding to the different batches, denoted with different point shapes. 
Within each cluster, smaller groups of cells of the same type are present; however, it is clear that the main clusters are driven by batch information instead of cell types.
Therefore, results from the upper panel of Figure~\ref{fig:sim1} do not allows us to properly identify regions of the space of partitions which are consistent with the factors of scientific interest. 
The bottom panels of Figure~\ref{fig:sim1} illustrate results for BC-$t$-SNE, Harmony and MNN and show that, after adjustment, the effect of unwanted batches is effectively removed from the $t$-SNE embeddings.
Indeed, different point shapes are uniformly spread across the four main clusters, which now correspond to the different cell types denoted with different colors.
From visual inspection, all the competitors achieve satisfactory results in terms of removing batch effects while preserving cell types, with BC-$t$-SNE highlighting the presence of different clusters more distinctly than the competitors.
Such preliminary findings are quantitatively evaluated in Table~\ref{tab:sim}, where the ability of the methods in removing batches and preserving cell types is evaluated in terms of silhouette coefficients, using the scone software \citep{cole2019performance}, kBET test metric \citep{buttner2019test}, average LISI score \citep{korsunsky2019fast} and PC regression using scater \citep{mccarthy2017scater}.
All the measures have been normalised and rescaled in $[0-1]$, with $0$ indicating perfect separation across groups and $1$ perfect integration; note that the interpretation of such metrics is different depending on the partitioning under investigation. 
Specifically, good performance in terms of batch effect removal corresponds to large values of the proposed metrics, while adequate preservation of cell types is associated with small values \citep[e.g.][]{korsunsky2019fast}.
Table~\ref{tab:sim} indicates that all the methods achieve good performance in terms of batch-removal, with BC-$t$-SNE being most accurate in terms of kBET, average LISI and highly competitive in terms of rescaled silhouette coefficient.
Coherency with Figure~\ref{fig:sim1}, the second half of Table~\ref{tab:sim} shows that BC-$t$-SNE outperforms the competitors in terms of conservation of cell types.

\begin{table}[ht]
	\caption{Simulation study. Evaluation of batch removal and cell types preservation. Best performance is reported in boldface.}
	\label{tab:sim}
\centering
\begin{tabular}{rrrrrrr}
	\toprule
 && SIL & kBET & LISI & PcReg \\
	\toprule
	\textbf{Batches}   & BC-$t$-SNE & 0.983          & \textbf{0.999} & \textbf{0.741} & 0.000 \\
	                   & Harmony    & 0.978          & 0.997          & 0.733          & 0.000 \\
  	                   & MNN        & \textbf{0.984} & 0.921          & 0.668          & 0.000 \\
			   \midrule
\textbf{Cell types}        & BC-$t$-SNE & \textbf{0.428} & \textbf{0.294} & \textbf{0.011} & 1.000 \\
	                   & Harmony    & 0.473          & 0.314          & 0.014          & 1.000 \\
  	                   & MNN        & 0.689          & 0.999          & 0.043          & 1.000 \\
			   \bottomrule
\end{tabular}
\end{table}
\section{Application}
\subsection{Dataset description}
Medulloblastoma is among the most frequent malignant brain tumors in children.
Recent studies have observed that the Sonic Hedgehog (SHH) signaling pathway is hyperactivated in $30\%$ of human medulloblastoma, therefore stimulating novel studies in this direction  \citep{zurawel2000analysis,ellison2011medulloblastoma}.
Activation of the SHH pathway, which stimulates proliferation of granule cell neurons during cerebellar development, has been used to create genetically engineered mice for scientific purposes, with the SmoM2 process being a routinely used pipeline \citep{rubin2006targeting}.
Specifically, SmoM2 mice have a transgenic mutated Smo allele which was originally isolated from a tumor and can be engineered to be not expressed until acted upon by Cre recombinase \citep{mao2006novel,helms2000autoregulation,machold2005math1}.
These mice are mated with genetically engineered matches that express Cre recombinase in cerebellar granular neuron progenitors, leading to descendants which develop medulloblastoma with $100\%$ frequency by postnatal day $12$.

Data used in this section come from $5$ mice at postnatal day $12$ created using such a pipeline and analysed under different sessions.
Specifically, mice $1,2$ (Females) on July 2\textsuperscript{nd}, mouse $3$ (Male) on July 25\textsuperscript{th} and mice $4,5$ (Males) on August 18\textsuperscript{th}.
Tumors were dissociated and individual cells co-encapsulated in a microfluidics chamber with primer-coated beads in oil-suspended droplets.
All primers on each bead contained a bead-specific bar code and an unique molecular identifier (UMI), followed by an oligo-dT sequence, while mRNAs were captured on the oligo-dT, reverse-transcribed and amplified.
Libraries were generated using the Drop-seq protocol V$3.1$ \citep{macosko2015highly}.
Following standard sequencing and preprocessing procedures, individual transcripts were identified by the UMI bar code, with cell identity inferred from the bead-specific bar codes. 
Analysis on the normalised data has been restricted to cells with more than $500$ detected genes. 
Furthermore, outlier cells with more than $4$ standard deviations above the median number of genes were excluded from the analysis, along with UMIs and mitochondrial content per cell in order to address the common problems of gene drop out, unintentional cell-cell multiplexing and premature cell lysis \citep{vladoiu2018childhood}.
The resulting pre-processed expression matrix $\X$ consists of $p=16680$ genes measured over $n=17746$ different cells; for each mouse, the number of valid cells was $3381$, $3402$, $1454$, $1647$ and $8062$, respectively.

The focus of our analysis is on evaluating if the proposed BC-$t$-SNE method provides a robust data representation, which is successful at removing batch effects without affecting biological information of interest; see \citet{ocasio2019scrna} for an analysis involving cell-annotations on the same dataset.

\begin{figure}
	\centerline{\includegraphics[width = .5\textwidth]{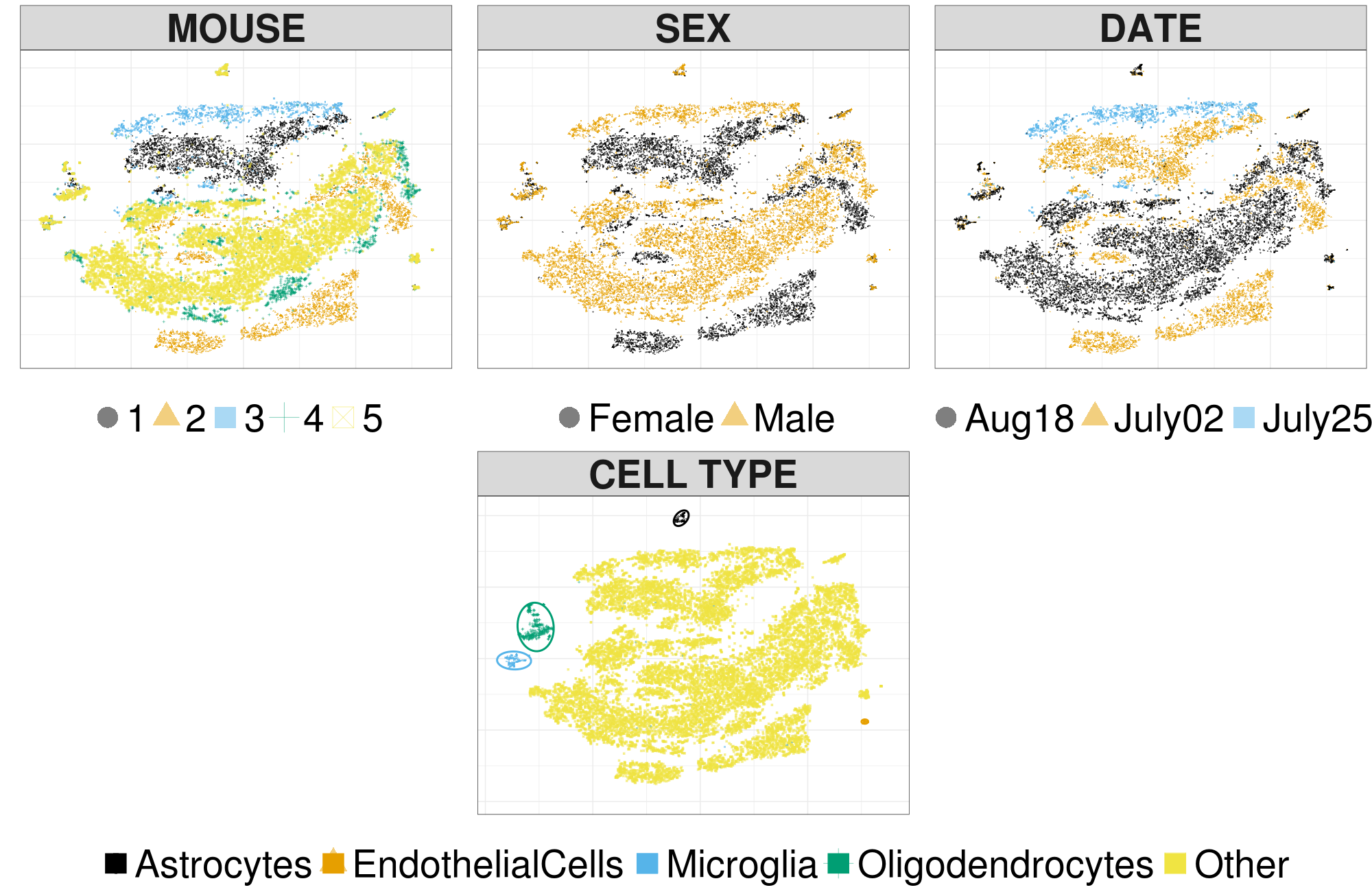}}
	\caption{Unadjusted $t$-SNE coordinates. Points and shapes vary with batches.} 
\label{fig2}
\end{figure}
\subsection{Results}
The presence of batch effects is investigated via unadjusted $t$-SNE embeddings, estimated over the first $k=50$ principal components; larger numbers of principal components resulted in less structured embeddings, and are not reported.
The first row of Figure~\ref{fig2} highlights systematic differences with batch membership, while the second row shows information on cell type.
Empirical results confirm the presence of strong batch effects, with respect to mouse identifier (first row, first column), sex of the mouse (first row, second column) and date of the experiment (first row, third column). 
For example, cells from mouse $1$ form a cluster which is clearly distinct from the others.
As expected, we observe some overlap between batch variables due to the experimental design.
The second row of Figure~\ref{fig2} highlights differences across cell types, confirming that unadjusted $t$-SNE
produces isolated clusters which are in agreement with the cell types indicated in \citet{ocasio2019scrna}.

\begin{figure}
	\centerline{\includegraphics[width = .5\textwidth]{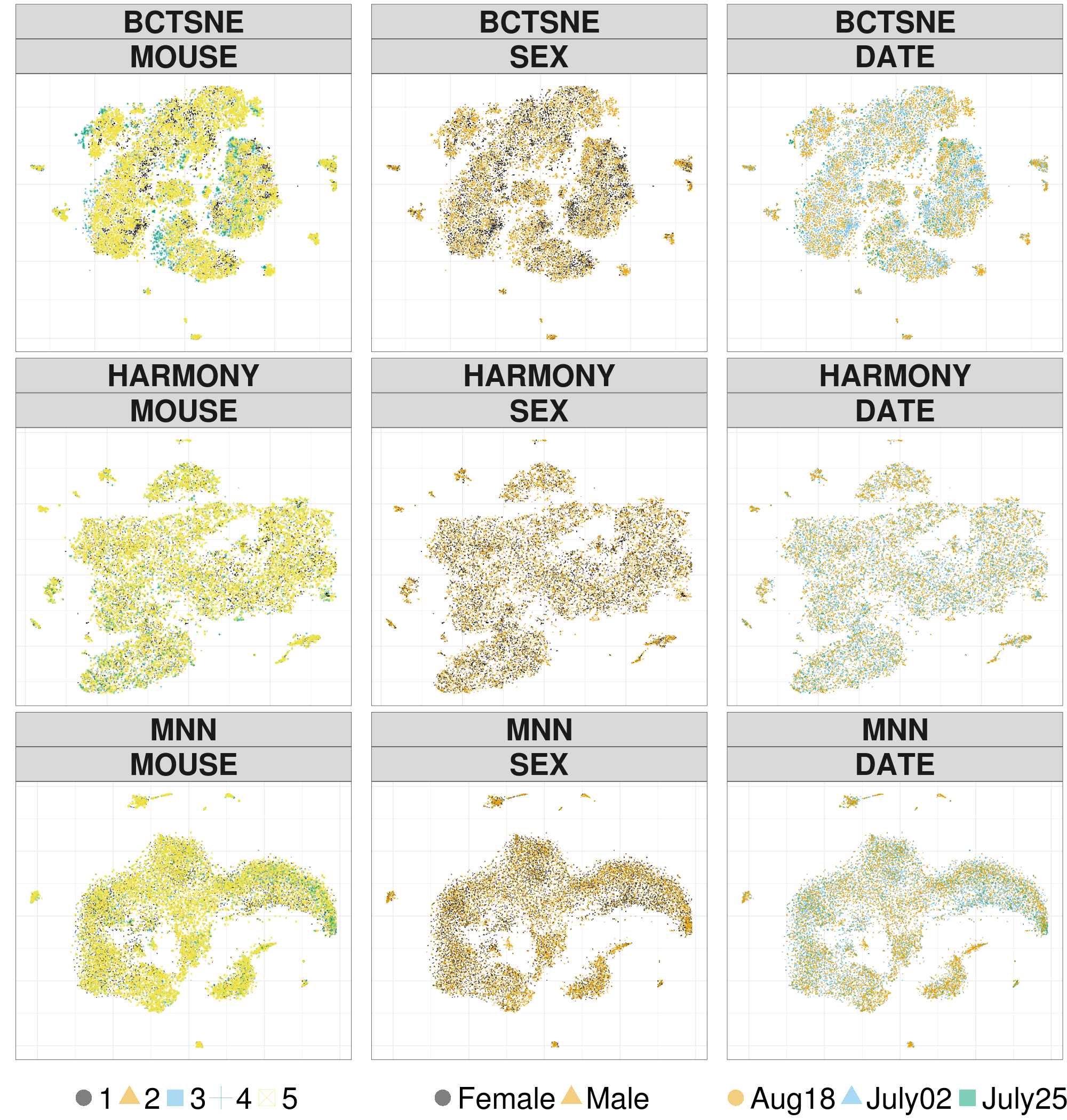}}
	\caption{$t$-SNE coordinates after correction. Points and shapes vary with batches.} 
\label{fig4}
\end{figure}

Figure~\ref{fig4} refers to adjusted $t$-SNE coordinates, estimated with Algorithm~\ref{alg:og} using the same settings described in the simulation study. 
We compare BC-$t$-SNE with the same approaches used in the simulation studies.
The first row of Figure~\ref{fig4} refers to BC-$t$-SNE; second and third to Harmony and MNN, respectively. Results suggest a satisfactory performance in terms of batch effect removal for all the methods considered.
Indeed, BC-$t$-SNE embeddings from Figure~\ref{fig4} 
show no evidence of systematic variation with any of the batch variables under investigation.
Such batches are marked by the color and shape of points in Figure~\ref{fig4}, showing that the batches are spread homogeneously across the embedded space after adjustment.

\begin{table}[t]
\centering
\caption{Evaluation of batch removal.}
	\label{tab:tab1}
\begin{tabular}{rlrrrrrr}
 && SIL & kBET & LISI & pcR \\
			 \midrule
	BC-$t$-SNE            & Sex   & 0.995 & 0.166 & 0.659& 1.000 \\
		              & Date  & 0.980 & 0.235 & 0.457& 1.000 \\
  		              & Mouse & 0.975 & 0.829 & 0.368& 1.000 \\
			 \midrule
	Harmony               & Sex   & 0.997 & 0.299 & 0.846& 1.000 \\
  		              & Date  & 0.987 & 0.421 & 0.540& 1.000 \\
  		              & Mouse & 0.975 & 0.854 & 0.498& 0.999 \\
			 \midrule
	MNN                   & Sex   & 0.999 & 0.219 & 0.844& 1.000 \\
  		              & Date  & 0.996 & 0.392 & 0.546& 1.000 \\
  		              & Mouse & 0.958 & 0.794 & 0.502& 0.999 \\
  \bottomrule
\end{tabular}
\end{table}

Following the metrics used in the simulations, Table~\ref{tab:tab1} quantitatively evaluates the success in removing batch effects.
Results indicate that BC-$t$-SNE achieves a performance which is highly competitive with the other approaches. 
Focusing, for example, on mouse identifiers, the normalised silhouette coefficient suggests that BC-$t$-SNE removes batches more effectively than MNN and Harmony; similar conclusions hold also when kBET is considered.
According to iLISI, instead, the baseline data adjustment methods perform better than BC-$t$-SNE.
This result is not surprising, since such approaches optimize objective functions which are directly related to the iLISI metric \citep{korsunsky2019fast}.

\begin{figure}
	\centerline{\includegraphics[width = .4\textwidth]{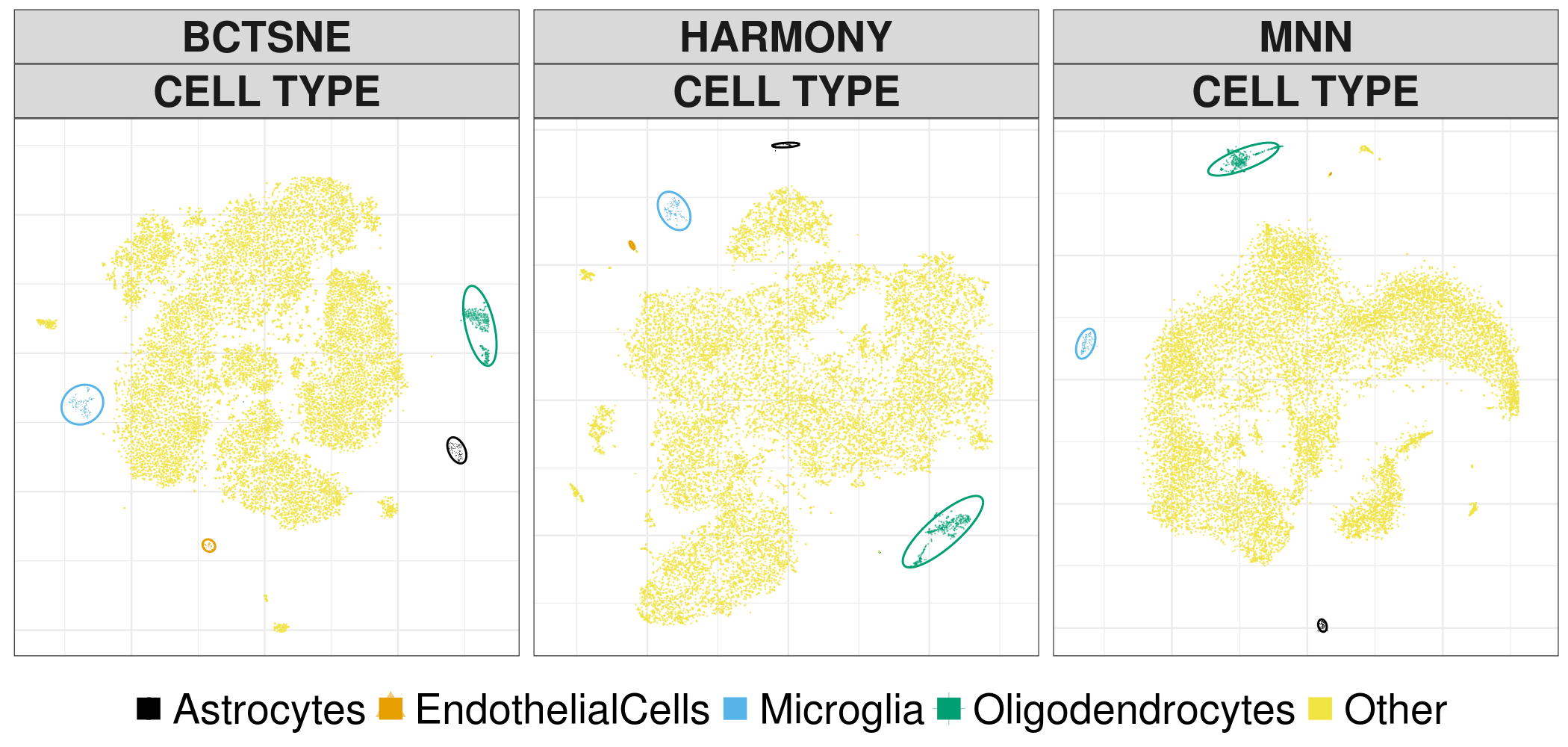}}
	\caption{$t$-SNE coordinates after adjustment. Points and shapes vary with cell types.} 
\label{fig5}
\end{figure}
\begin{table}[t]
\centering
\caption{Evaluation of cell-type preservation.}
	\label{tab:tab2}
\begin{tabular}{rlrrrrrr}
 & SIL & kBET & iLISI & pcR \\
  \hline
BC-$t$-SNE & 0.052 & 0.000 & 0.000 & 0.000 \\
  MNN      & 0.056 & 0.000 & 0.000 & 0.000 \\
  Harmony  & 0.031 & 0.000 & 0.000 & 0.000 \\
   \hline
\end{tabular}
\end{table}
Lastly, it is important to investigate that after removing batch effects, clusters of cell types associated with medulloblastoma are preserved in the low-dimensional coordinates, and similar cells are close in the embedded space.  Figure~\ref{fig5} shows results for BC-$t$-SNE and the baseline methods. 
The bulk of cells in a large central cluster are from tumors having markers within in a range of differentiation states, ranging from proliferative, undifferentiated cells expressing the SHH-pathway transcription factor Gli1, to cells in successive states of CGN differentiation, marked by sequential expression of markers Ccnd2, Barhl1, Cntn2, Rbfox3, and Grin2b \citep{ocasio2019scrna}.
Clusters of cells surrounded with colored ellipses correspond to endothelial cells,  microglia,  oligodendrocytes and astrocytes, which are common in the stroma within or adjacent to the tumors. 
Empirical findings suggest that such clusters are correctly preserved after adjustment; see also Table~\ref{tab:tab2} for a quantitative evaluation.

\section{Discussion}
In this article we have introduced BC-$t$-SNE, a novel modification of $t$-SNE which allows for correction for multiple batch effects during estimation of the low-dimensional embeddings.
The proposed approach has demonstrated good performance in simulation studies and on an application involving mouse medulloblastoma, where unwanted variations are successfully removed without
removing information on differences across cell types.

A possible extension for future development involves adapting the proposed procedure to more efficient optimisation of the $t$-SNE loss function, to overcome the computational constraints encountered with large $n$ \citep{van2014accelerating}.
One way to address such an issue involves the use of alternative gradient methods, with methods based on stochastic gradients being popular in the literature. 


\section*{Funding}
The work of Emanuele Aliverti and David B. Dunson was partially funded by the grant ``Fair predictive modelling'' from the Laura \& John Arnold Foundation.
We thank the UNC CGBID Histology Core supported by P30 DK 034987, the UNC Tissue Pathology Laboratory Core supported by NCI CA016086 and UNC UCRF, and the UNC Neuroscience Center Confocal and Multiphoton Imaging and bioinformatics cores supported by The Eunice Kennedy Shriver National Institute of Child Health and Human Development (U54HD079124) and NINDS (P30NS045892). J.O. was supported by NINDS (F31NS100489). 
T.R.G. was supported by NINDS (R01NS088219, R01NS102627, R01NS106227) and by the UNC Department of Neurology Research Fund. T.R.G., K.W., and B.B. were supported by a TTSA grant from the NCTRACS Institute, which is supported by the National Center for Advancing Translational Sciences (NCATS), National Institutes of Health, through Grant Award Number UL1TR002489. 

\bibliographystyle{apalike}
\bibliography{99BIB.bib}

\end{document}

%% file: pack.tex
\usepackage{amsmath,amsthm,amssymb,mathabx,mathtools}
\usepackage{graphicx,booktabs,xcolor}
\usepackage[top = 1.8cm , bottom= 2cm , left= 2cm , right= 2cm]{geometry}
\usepackage[colorlinks,citecolor=bl]{hyperref}

\definecolor{prp}{HTML}{6C71C4}
\definecolor{bl}{HTML}{0076A1}
\usepackage{algorithmicx}
\usepackage{algorithm2e}
\usepackage{setspace}
\usepackage[shortlabels]{enumitem}

\usepackage{mdframed}
\mdfdefinestyle{st}{%
    linecolor=bl,
   linewidth=2pt,
        }

\usepackage{titlesec}
\titleformat{\section}
  {\centering\bfseries}
  {\thesection}
  {.5em}
  {\MakeUppercase}

\titleformat{\subsection}
  {\normalfont\bfseries\itshape}{\thesubsection}{1em}{}

\usepackage[noabbrev,capitalise]{cleveref}

